\documentclass[11pt]{article}
\usepackage{moriond,epsfig}

\newcommand\micron{\mu\mathrm{m}}

\newcommand{\jpsi}{J/$\psi$}

\begin{document}
\vspace*{4cm}
\title{Study of dimuon production in Indium-Indium collisions\\
with the NA60 experiment}

\author{R.~Shahoyan$^{2,5}$,
R.~Arnaldi$^{9}$,
K.~Banicz$^{2,4}$,
J.~Castor$^{3}$,
B.~Chaurand$^{7}$,
C.~Cicalo$^{1}$,
A.~Colla$^{9}$,
P.~Cortese$^{9}$,
S.~Damjanovic$^{4}$,
A.~David$^{2,5}$,
A.~de~Falco$^{1}$,
A.~Devaux$^{3}$,
L.~Ducroux$^{6}$,
H.~En'yo$^{8}$,
A.~Ferretti$^{9}$,
M.~Floris$^{1}$,
P.~Force$^{3}$,
N.~Guettet$^{2,3}$,
A.~Guichard$^{6}$,
H.~Gulkanian$^{10}$,
J.~Heuser$^{8}$,
M.~Keil$^{2,4}$,
L.~Kluberg$^{2,7}$,
J.~Lozano$^{5}$,
C.~Louren\c{c}o$^{2}$,
F.~Manso$^{3}$,
A.~Masoni$^{1}$,
P.~Martins$^{2,5}$,
A.~Neves$^{5}$,
H.~Ohnishi$^{8}$,
C.~Oppedisano$^{9}$,
P.~Parracho$^{2}$,
P.~Pillot$^{6}$,
G.~Puddu$^{1}$,
E.~Radermacher$^{2}$,
P.~Ramalhete$^{2,5}$,
P.~Rosinsky$^{2}$,
E.~Scomparin$^{9}$,
J.~Seixas$^{2,5}$,
S.~Serci$^{1}$,
P.~Sonderegger$^{5}$,
H.J.~Specht$^{4}$,
R.~Tieulent$^{6}$,
G.~Usai$^{1}$,
R.~Veenhof$^{2,5}$ and
H.K.~W\"ohri$^{2,5}$}

~

\address{
$^{~1}$Univ.\ di Cagliari and INFN, Cagliari, Italy;
$^{~2}$CERN, Geneva, Switzerland;
$^{~3}$LPC, Univ.\ Blaise Pascal and CNRS-IN2P3, Clermont-Ferrand, France;
$^{~4}$Univ.\ Heidelberg, Heidelberg, Germany;
$^{~5}$IST-CFTP, Lisbon, Portugal;
$^{~6}$IPN-Lyon, Univ.\ Claude Bernard Lyon-I and CNRS-IN2P3, Lyon, France;
$^{~7}$LLR, Ecole Polytechnique and CNRS-IN2P3, Palaiseau, France;
$^{~8}$RIKEN, Wako, Saitama, Japan;
$^{~9}$Univ.\ di Torino and INFN, Italy;
$^{10}$YerPhI, Yerevan, Armenia}

\maketitle\abstracts{The NA60 experiment at the CERN-SPS is devoted to
the study of dimuon production in heavy-ion and proton-nucleus
collisions. We present preliminary results from the analysis of
Indium-Indium collisions at 158~GeV per nucleon.  The topics covered
are low mass vector meson production, J/$\psi$ production and
suppression, and the feasibility of the open charm measurement from
the dimuon continuum in the mass range below the J/$\psi$ peak.}


The study of dimuon production in heavy-ion collisions is generally
considered to be one of the most powerful tools in the search for the
phase transition between the normal nuclear matter and the Quark-Gluon
Plasma phase, where the quarks and gluons are no longer confined into
hadrons. The most intriguing findings of dilepton experiments working
in this field have been: the excess in the production of dielectron
pairs in the mass window 200--700~MeV/$c^2$, together with the
flattening of the $\rho$ and $\omega$ peaks, observed by the
NA45/CERES experiment in \mbox{S-Au} and \mbox{Pb-Au}
collisions~\cite{CERESLMR}; the anomalous J/$\psi$ suppression found
by NA50 in central \mbox{Pb-Pb} collisions~\cite{NA50Psi}; and the
excess in the production of dimuons in the ``intermediate mass
region'', 1.2--2.7~GeV/$c^2$, seen by NA38 and NA50 in \mbox{S-U} and
\mbox{Pb-Pb} collisions~\cite{NA50IMR}, and by HELIOS-3 in S-W
collisions~\cite{Helios3}.  This paper presents the analysis of these
topics using $\sim$\,230 million Indium-Indium dimuon events collected
by NA60 in 2003.

\begin{figure}[ht]
\centering
\begin{tabular}{cc}
\begin{minipage}{0.45\textwidth}
\resizebox{0.99\textwidth}{!}{%
  \includegraphics*{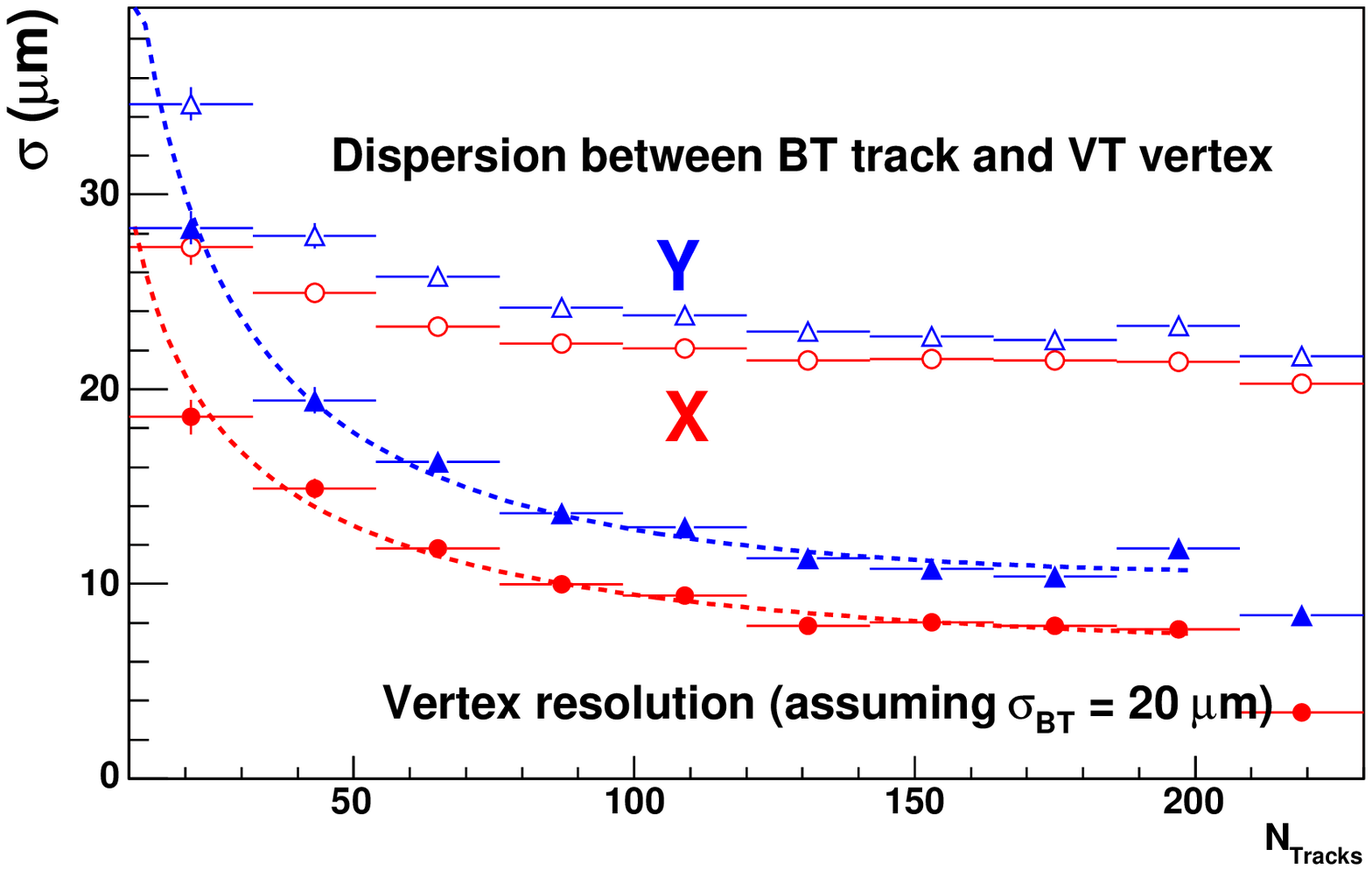}}
\vglue -3mm
\caption{Dispersion between the fitted vertex and the Beam Tracker
prediction (open symbols) and extracted vertex resolution (closed
symbols), assuming a Beam Tracker precision of 20~$\micron$.}
\label{fig:vtbt}
\end{minipage}
&
\begin{minipage}{0.45\textwidth}
\resizebox{0.93\textwidth}{!}{%
  \includegraphics*{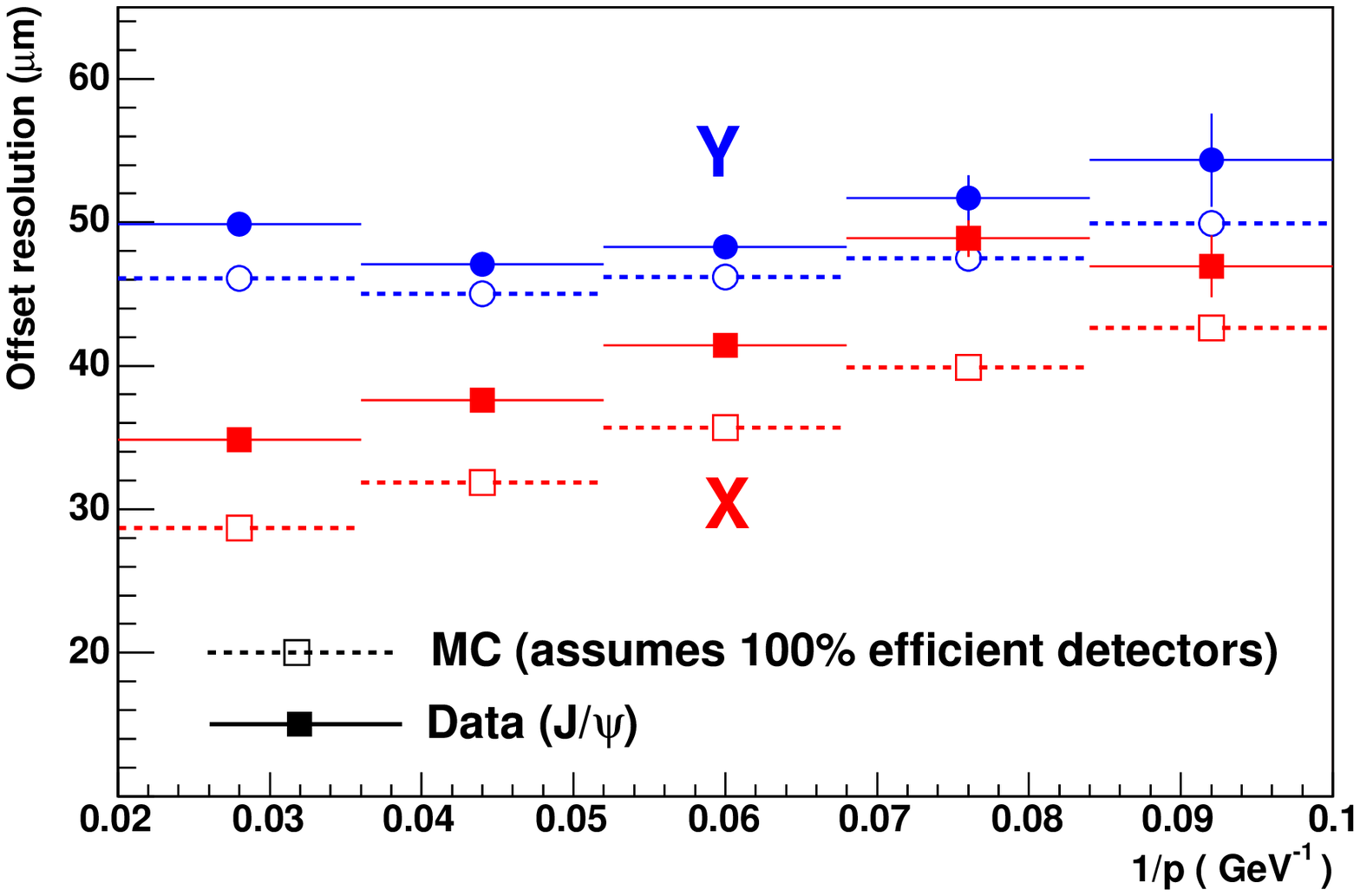}}
\vglue -3mm
\caption{Measured offset resolution for J/$\psi$ muons (solid symbols)
compared to the resolution expected from a Monte Carlo simulation,
assuming fully efficient pixel planes (open symbols).}
\label{fig:offsres}
\end{minipage}
\end{tabular}
\end{figure}

The NA60 apparatus complements a Muon Spectrometer and a Zero Degree
Calorimeter with a Vertex Telescope~\cite{refApparatus} made of
Silicon pixel planes embedded in a 2.5~T dipole field and a Beam
Tracker measuring the transverse coordinates of the incoming ion
before its interaction in the target.  The Vertex Telescope allows us
to reconstruct the interaction vertex.  Figure~\ref{fig:vtbt} shows
the dispersion between the transverse position of the fitted vertex
and of the Beam Tracker prediction, as a function of the number of
tracks associated with the vertex. The corresponding vertex resolution
is also shown. For most of our data it is better than 10~$\micron$ in
$X$ (bending plane) and better than 15~$\micron$ in $Y$.

The essential feature of NA60 is the matching between the muons
reconstructed in the Muon Spectrometer and the tracks measured in the
Vertex Telescope before they scatter in the hadron absorber. This
improves the mass resolution from $\sim$\,75~MeV/$c^2$ to
$\sim$\,20~MeV/$c^2$, at the $\omega$ mass. Additionally, the
measurement of the impact parameter (offset) of the muons at the
production vertex allows us to tag the muons from open charm decays.
Figure~\ref{fig:offsres} shows the offset resolution measured for
J/$\psi$ muons, which are expected to come exactly from the
interaction point, as a function of their inverse momentum. Expected
values from Monte Carlo simulations are also shown.  In order to take
into account the dependence of the muon offset resolution on its
momentum, the measured offsets are weighted by the inverse of their
covariant error matrices, $V^{-1}$, which incorporate the
uncertainties from the vertex fit and from the muon extrapolation,
$\Delta = \sqrt{\Delta x^2 V_{xx}^{-1} + \Delta y^2 V_{yy}^{-1} + 2
\Delta x \Delta y V_{xy}^{-1}}$.  Another benefit of the matching
procedure is a strong reduction of the combinatorial background from
uncorrelated $\pi$ and K decays.


\begin{figure}[ht]
\centering
\begin{tabular}{cc}
\begin{minipage}{0.45\textwidth}
\resizebox{0.99\textwidth}{55mm}{%
  \includegraphics*{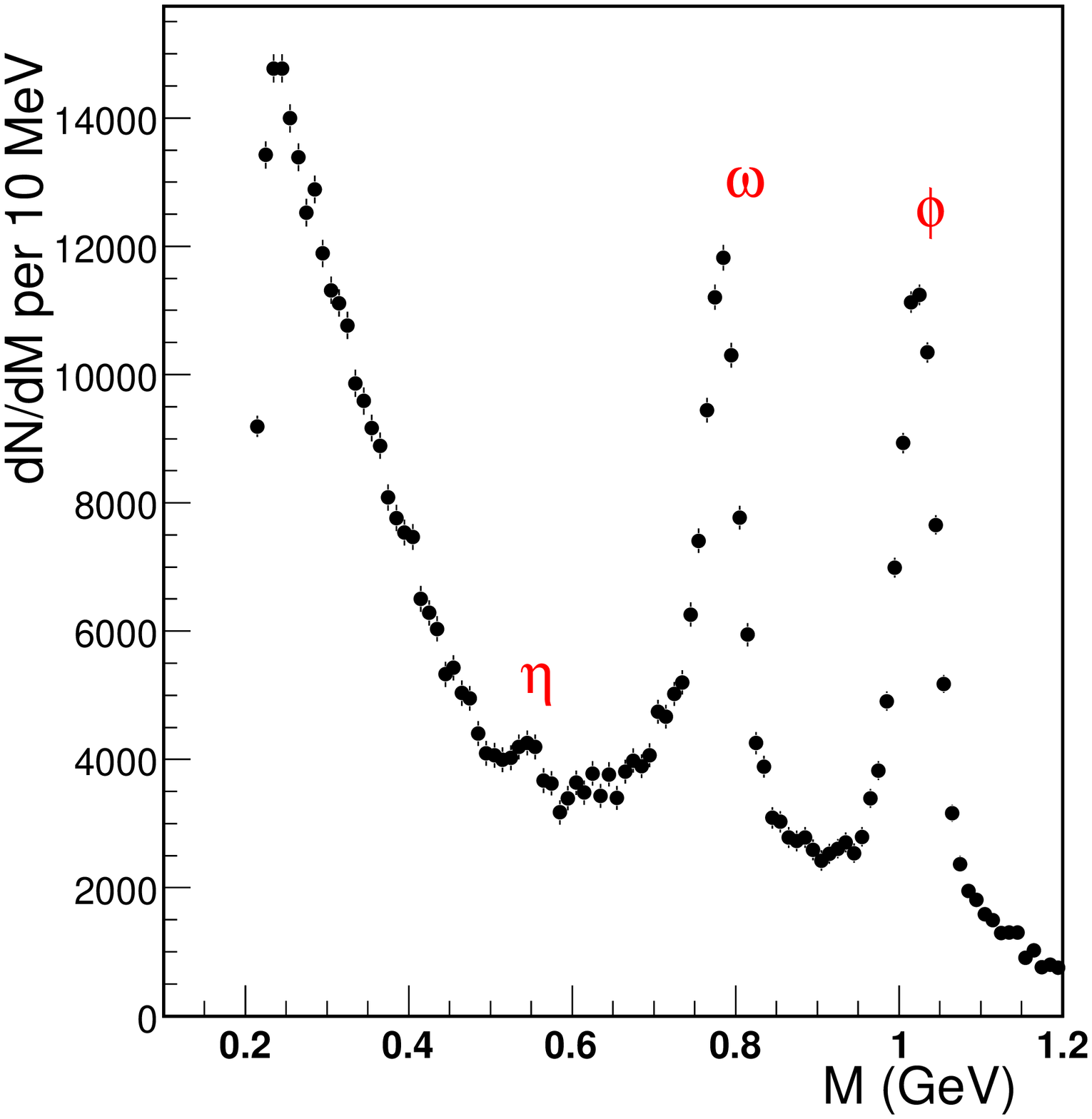}}
\vglue -4mm
\caption{Dimuon mass distribution in In-In collisions, after background
subtraction.}
\label{fig:dimuon-mass}
\end{minipage}
&
\begin{minipage}{0.45\textwidth}
\resizebox{0.99\textwidth}{53mm}{%
  \includegraphics*{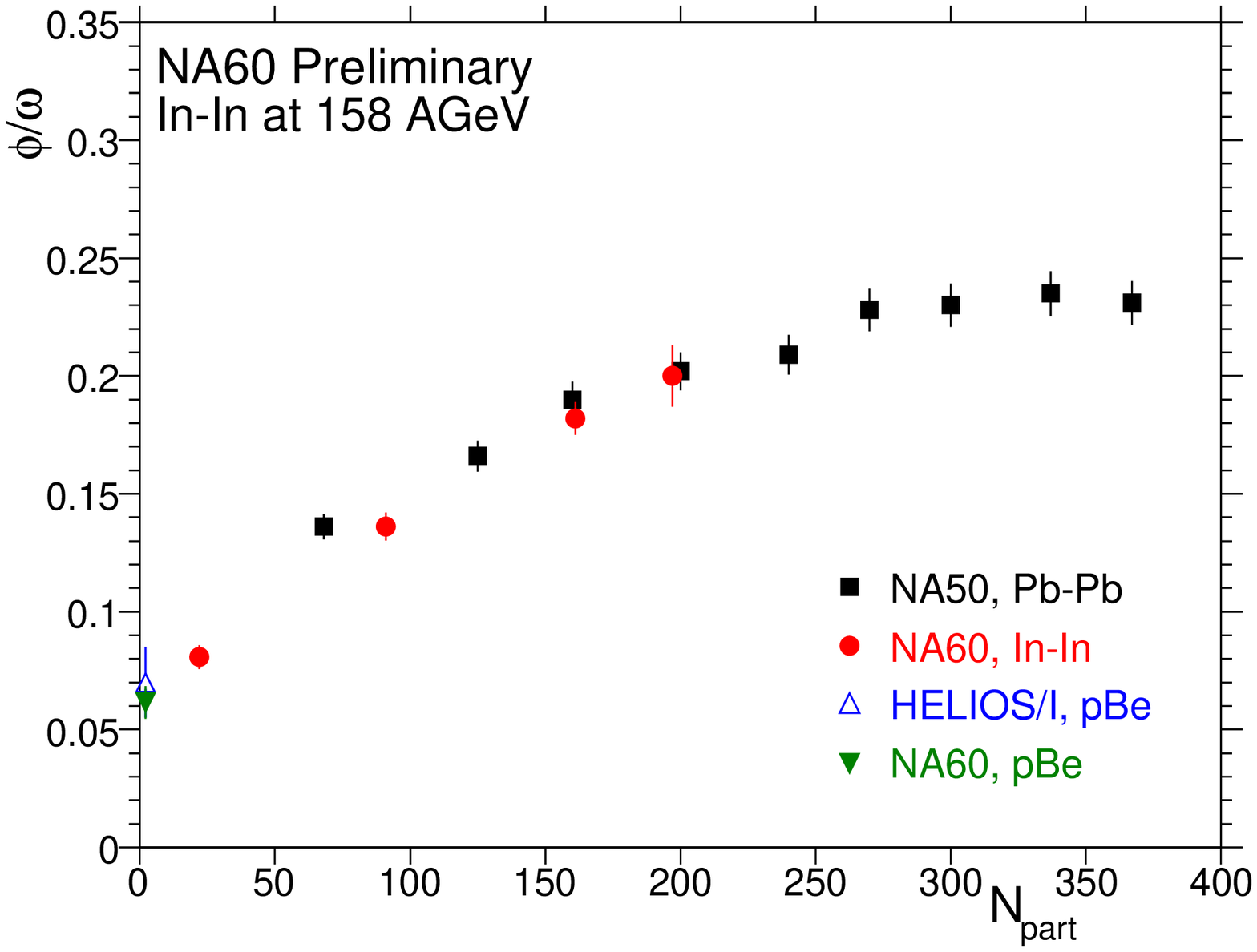}}
\vglue -4mm
\caption{The $\phi/\omega$ cross section ratio as a function of the
  number of participants.}
\label{fig:phiomega}
\end{minipage}
\end{tabular}
\end{figure}
\begin{figure}[ht]
\centering
\vglue -2mm
\begin{tabular}{c}
\resizebox{0.49\textwidth}{50mm}{%
\includegraphics*{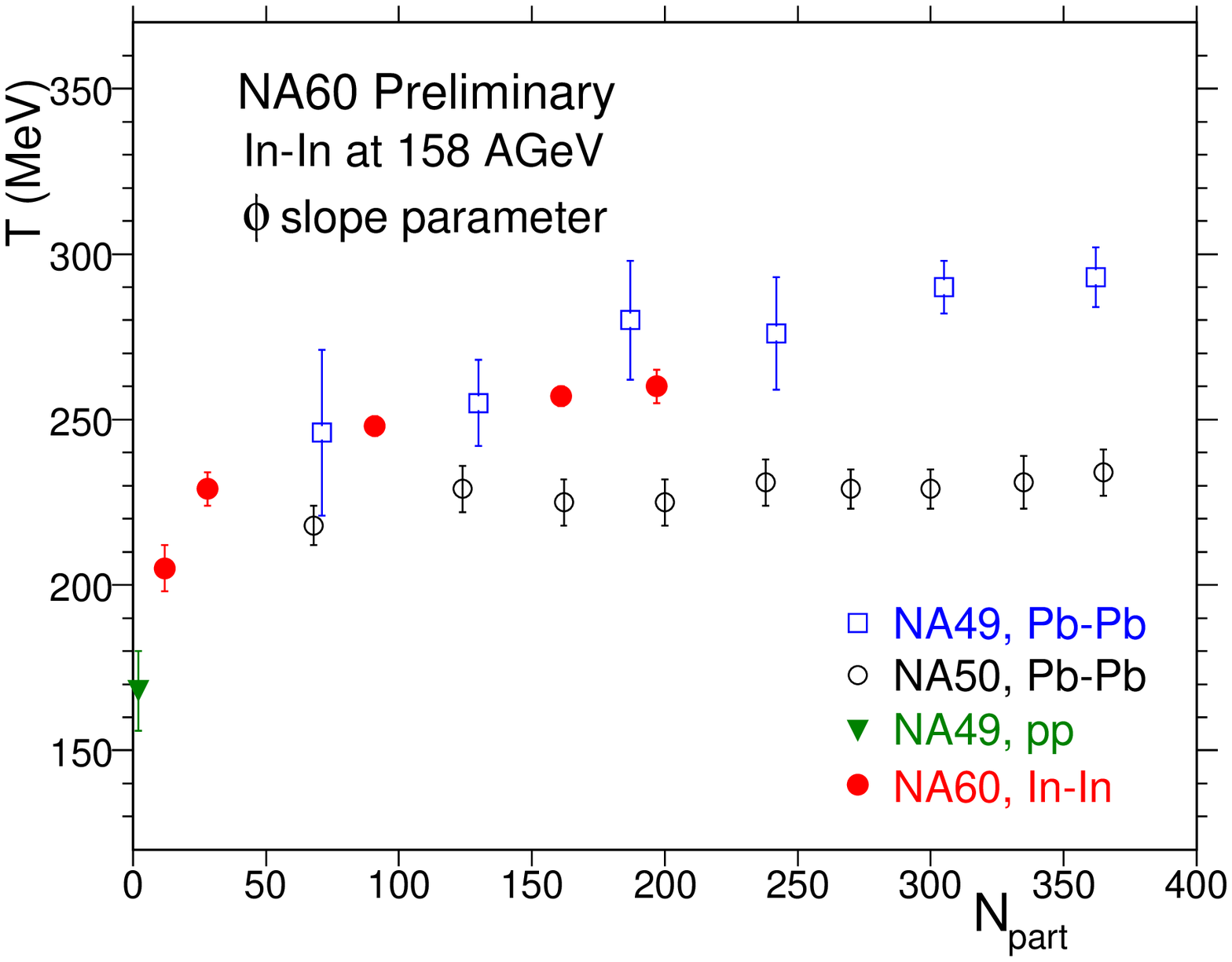}}
\resizebox{0.49\textwidth}{50mm}{%
\includegraphics*{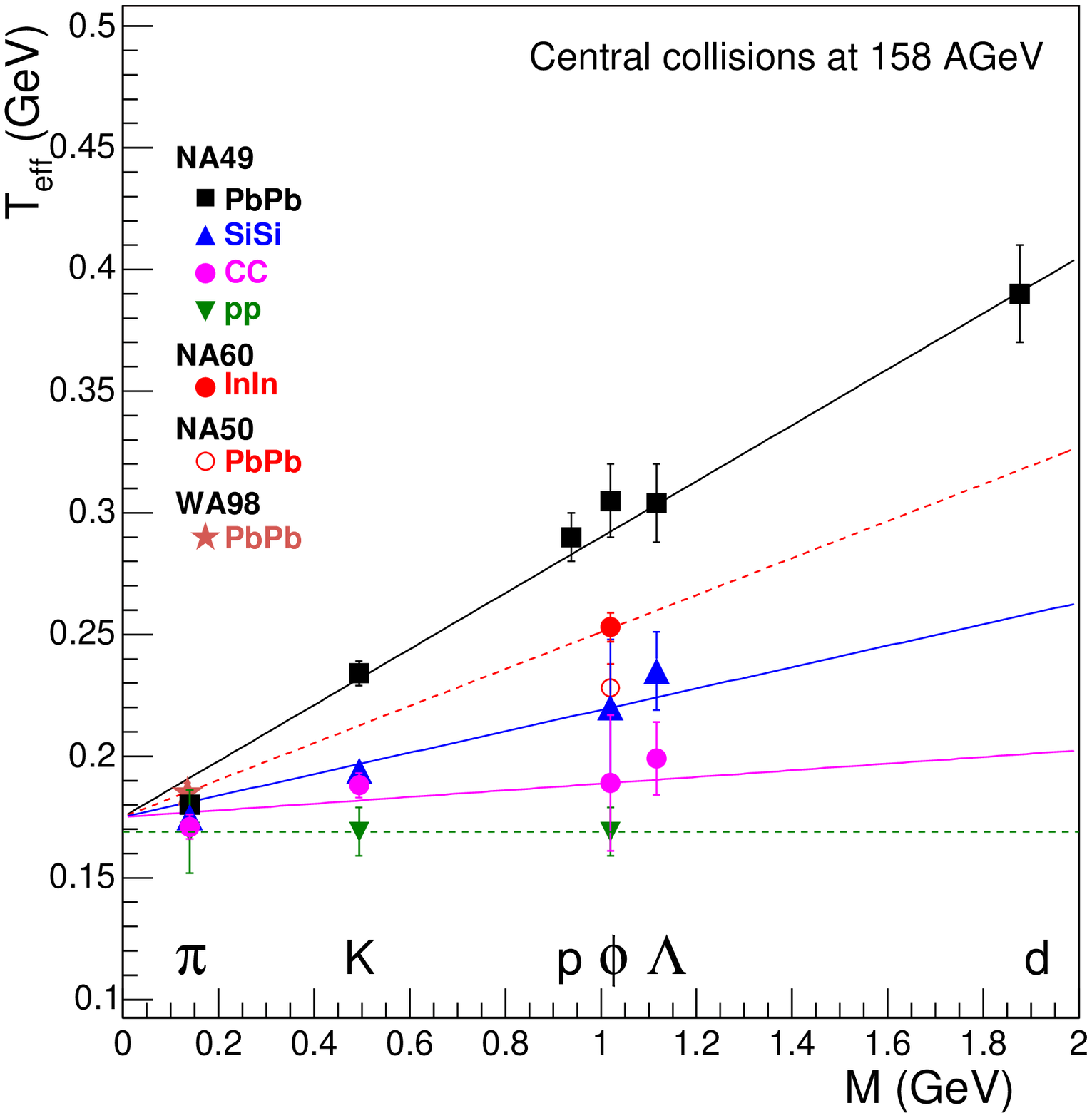}}
\end{tabular}
\vglue -3mm
\caption{The $\phi$ $m_{\rm T}$ inverse slope parameter, $T$, as a
function of the number of participants (left) and of the particle mass
and colliding system (right).}
\label{fig:phi-T}
\end{figure}

Figure~\ref{fig:dimuon-mass} shows the dimuon mass distribution
measured in In-In collisions, after combinatorial background
subtraction, integrating all collision centralities. The mass
resolution is 23~MeV at the $\phi$.  The extracted $\phi/\omega$
values are shown in Fig.~\ref{fig:phiomega}, in four centrality bins,
as a function of the number of participants evaluated from a Glauber
fit to the energy spectrum measured by the ZDC.  The figure includes
values obtained from the NA50 measurements in Pb-Pb
collisions~\cite{na50phi}.  We have assumed that the $\rho$ and
$\omega$ production cross sections are identical.  In order to have
both data sets reported in the same phase space region, we converted
the NA50 points from the window $m_{\rm T}>1.5$~GeV to the window
$p_{\rm T}>1.1$~GeV/$c$, using the inverse slope parameter measured by
NA50, $T=228$~MeV, for the extrapolation~\cite{GU-HP04}.  The Pb-Pb
and In-In values are in good agreement, suggesting that $N_{\rm part}$
is a good scaling variable to describe the $\phi/\omega$ ratio.  We
have studied the $\phi$ transverse momentum distribution, selecting
signal dimuons in a narrow window at the $\phi$ mass.  Thanks to the
presence of the dipole field in the vertex region we have a good
dimuon acceptance down to zero $p_{\rm T}$.
Figure~\ref{fig:phi-T}-left shows how the $m_{\rm T}$ inverse slopes
extracted from our data compare with the Pb-Pb values measured by
NA50~\cite{na50phi} and NA49~\cite{na49}, as a function of the number
of participants.  There is a clear increase of $T$ between peripheral
and central In-In collisions, from $\sim$\,218~MeV to $\sim$\,255~MeV,
with $T=252\pm3$~MeV when integrating over all collision centralities.
Figure~\ref{fig:phi-T}-right compares our extracted $\phi$ temperature
with the measurements of other experiments as a function of the
particle mass and colliding system.

\begin{figure}[ht]
\centering
\vglue -2mm
\begin{tabular}{c}
\resizebox{0.49\textwidth}{!}{%
\includegraphics*{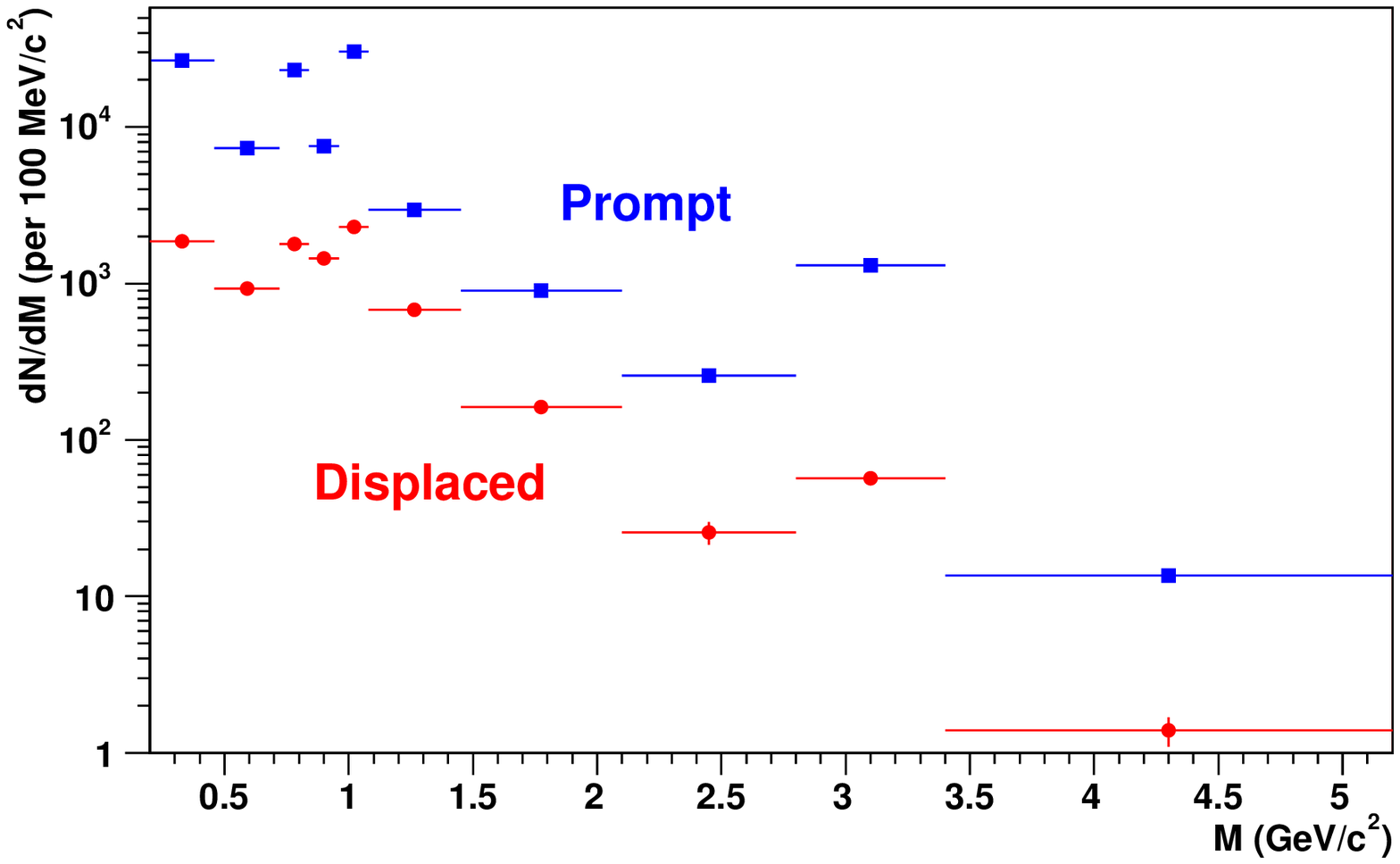}}
\resizebox{0.49\textwidth}{!}{%
\includegraphics*{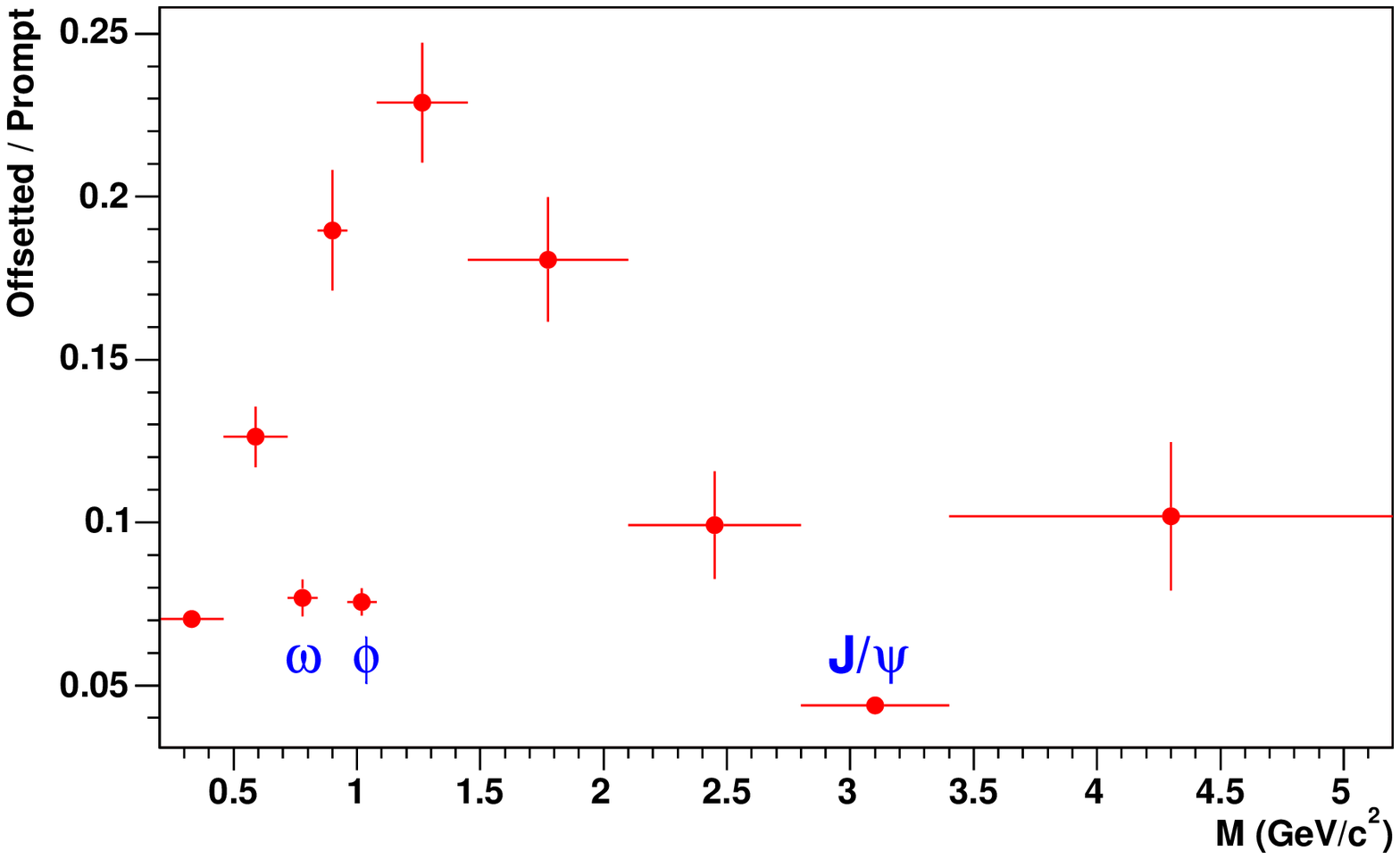}}
\end{tabular}
\vglue -3mm
\caption{Dimuon mass spectra from the ``displaced'' and ``prompt''
  samples (left), and their ratio (right).}
\label{fig:off_prompt}
\end{figure}

We tag the dimuons as being ``displaced'' if both muons have weighted
offsets above~1 (roughly 90~$\micron$ for muons with momenta around
15~GeV/$c$), otherwise they are tagged as ``prompt''.  To reject
events where the interaction vertex was misidentified, this selection
is only validated if the weighted transverse distance between the two
muons, at the $Z$ of the interaction vertex, is more than 0.7 for the
``displaced'' events and less than 2 for the ``prompt'' ones.  The
mass spectra for the two event samples are shown in
Fig.~\ref{fig:off_prompt}.  The regions dominated by the $\omega$,
$\phi$ and J/$\psi$ peaks are strongly suppressed in the displaced
sample, as is clearly shown in the ratio between displaced and prompt
mass spectra.

Among the interesting observations made by previous experiments
studying high-energy heavy-ion physics at the CERN SPS stands the
observation, by NA50, that the production yield of J/$\psi$ mesons is
suppressed in central Pb-Pb collisions beyond the normal nuclear
absorption defined by proton-nucleus data. This behaviour is expected
to occur if the matter produced in these extreme collisions goes
through a phase of deconfined partonic matter, where the charmonium
states should be dissolved when critical thresholds are exceeded,
either in the medium temperature (thermal transition, QGP) on in the
density of interacting partons (geometrical transition, percolation).
By comparing the centrality dependence of the suppression pattern
between two different colliding systems, Pb-Pb and In-In, we should be
able to identify the corresponding scaling variable and the physics
mechanism driving the suppression.

\begin{figure}[ht]
\centering
\vglue -2mm
\begin{tabular}{c}
\resizebox{0.49\textwidth}{55mm}{%
\includegraphics*{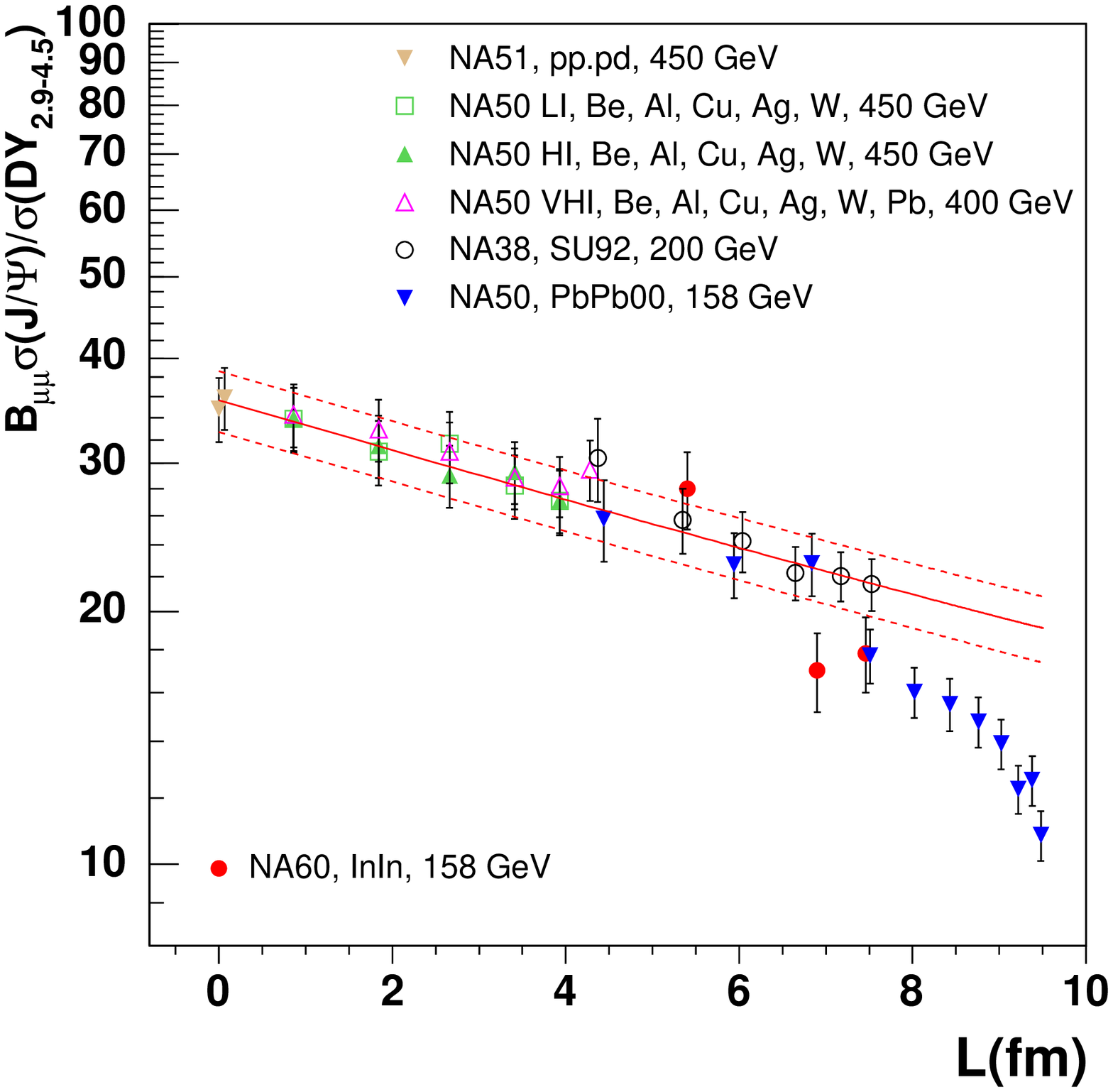}}
\resizebox{0.49\textwidth}{55mm}{%
\includegraphics*{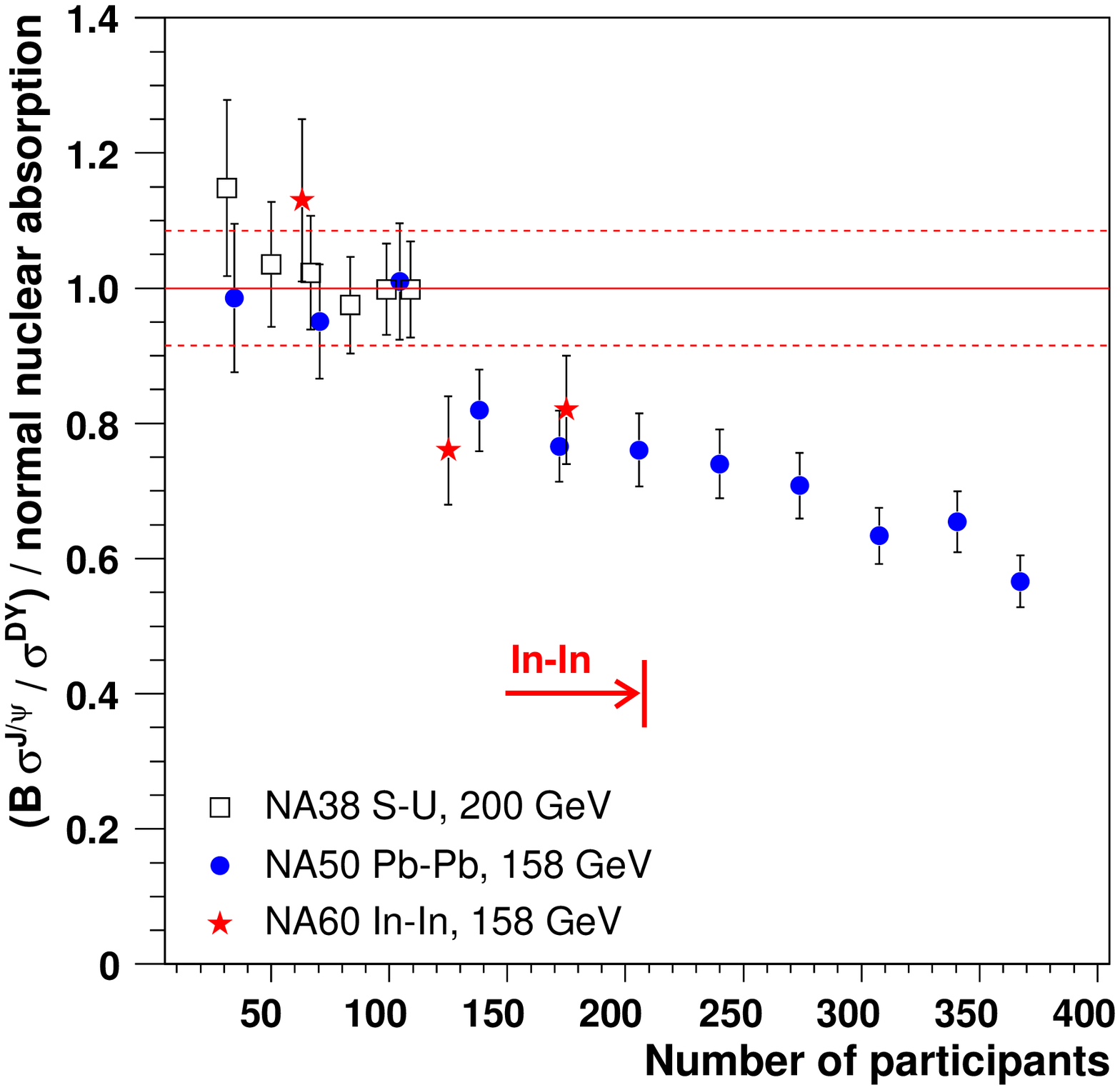}}
\end{tabular}
\vglue -3mm
\caption{\jpsi\ suppression before (left, versus $L$) and after
  (right, versus $N_{\rm part}$) dividing by the normal nuclear
  absorption curve.}
\label{fig:psisupp}
\end{figure}

Figure~\ref{fig:psisupp} shows the ratio between the J/$\psi$ and the
Drell-Yan production cross-sections measured in In-In collisions, in
three centrality bins, either as a function of $L$ (the distance of
nuclear matter crossed by the J/$\psi$ mesons after production) or
$N_{\rm part}$.  On the right panel the \jpsi\ suppression pattern is
divided by the normal nuclear absorption curve, defined by p-nucleus
data~\cite{GBorges}.  The J/$\psi$ and Drell-Yan cross-sections are
evaluated in the phase space window $2.92 < y_{\rm lab} < 3.92$ and
$-0.5 < \cos\theta_{\rm CS} <0.5$, where $\theta_{\rm CS}$ is the
polar decay angle of the muons in the Collins-Soper reference system.
The Drell-Yan value is given in the 2.9--4.5~GeV/$c^2$ mass window.
We see that, unlike what happens in the S-U collisions studied by
NA38, the J/$\psi$ production is suppressed in indium-indium
collisions beyond the normal nuclear absorption.  When the J/$\psi$
over Drell-Yan ratio is plotted as a function of $N_{\rm part}$ the
indium data points are in good agreement with the suppression pattern
measured in Pb-Pb.  The two sets of data points \emph{do not} overlap
as a function of $L$.  We are presently studying the J/$\psi$
suppression pattern without making the ratio with respect to the
Drell-Yan yield, so that we can have more centrality bins, and as a
function of the energy density, to clarify the origin of the anomalous
J/$\psi$ suppression.


We are also analysing the proton-nucleus data collected in 2004, with
158~GeV protons colliding on 7 different nuclear targets, to determine
the normal nuclear absorption of the \jpsi\ in the energy and
kinematical domains of the heavy-ion data, without the model dependent
assumptions presently used to correct the NA50 proton data, collected
at 400 or 450~GeV.

\medskip
This work was partially supported by the Funda\c{c}\~ao para a
Ci\^encia e a Tecnologia, Portugal.


\begin{thebibliography}{9}

\bibitem{CERESLMR} G.~Agakichiev {\em et al.} (CERES Coll.),
Phys. Rev. Lett. \textbf{75}, (1995) 1272; Phys. Lett. \textbf{B422}
(1998) 405; B.~Lenkeit {\em et al.} (CERES Coll.),
Nucl. Phys. \textbf{A661}, (1999) 23c.

\bibitem{NA50Psi} B.~Alessandro {\em et al.} (NA50 Coll.),
Eur. Phys. J. \textbf{C39} (2005) 335; A.~Baldit {\em et al.} (NA50
Coll.), Phys. Lett. \textbf{B140} (1997) 337.

\bibitem{NA50IMR} M.C.~Abreu {\em et al.} (NA50 Coll.),
Eur. Phys. J. \textbf{C14} (2000) 443.

\bibitem{Helios3} A.L.S. Angelis {\em et al.} (HELIOS-3 Coll.),
  Eur. Phys. J. \textbf{C13} (2000) 433.

\bibitem{refApparatus} K. Banicz {\it et al.},
Nucl. Instrum. Meth. \textbf{A539} (2005) 137.

\bibitem{na50phi} D. Jouan {\it et al.} (NA50 Coll.), J. Phys. {\bf
  G30} (2004) S277; B. Alessandro {\it et al.} (NA50 Coll.),
  Phys. Lett.  \textbf{B 555} (2003) 147.

\bibitem{GU-HP04} G. Usai \emph{et al.} (NA60 Coll.), Proc. of Hard
Probes 2004, Eur. Phys. J. \textbf{C}.

\bibitem{na49} V. Friese {\it et al.} (NA49 Coll.), Nucl. Phys. {\bf
  A698} (2002) 487c.

\bibitem{GBorges} G. Borges {\it et al.} (NA50 Coll.), these proceedings.

\end{thebibliography}
\end{document}